\documentclass[12pt]{article}

\input{epsf}

\begin{document}

\title{On the mean field treatment of attractive interactions in nonuniform simple
fluids}
\author{Kirill Katsov \\
\textit{Department of Physics, University of Washington}\\
\textit{P.O. Box 351560, Seattle, WA 98195-1560}\\
\\
John D. Weeks\\
\textit{Institute for Physical Science and Technology, and}\\
\textit{Department of Chemistry and Biochemistry}\\
\textit{University of Maryland, College Park, MD 20742}}
\date{}
\maketitle

\begin{abstract}
We study thermodynamic and structural properties of a Lennard-Jones liquid
at a state very close to the triple point as the radius of a hard sphere
solute is varied. Oscillatory profiles arise for small, molecular sized
radii while for large radii smooth interfaces with a ``drying layer'' of low
vapor density near the solute are seen. We develop a quantitative theory for
this process using a new mean field treatment where the effects of
attractive interactions are described in terms of a self-consistently chosen
effective single particle field. We modify the usual simple molecular field
approximation for the effective field in a very natural way so that exact
results (consistent with a given accurate equation of state for the uniform
fluid) arise in the ``hydrostatic limit'' of very slowly varying interfaces.
Very good agreement with the results of computer simulations for a wide
range of solute radii are found.
\end{abstract}

\section{Introduction}

In this paper we are concerned with the interplay between oscillatory
excluded volume correlations and the formation of smooth liquid-vapor
interfaces in simple liquids. Broadly speaking, the former arise because the
harshly repulsive molecular cores cannot overlap while the latter are
associated with the longer ranged and more slowly varying attractive
intermolecular interactions. The theoretical challenge is to develop a
general approach that can naturally explain the complete range of behavior.
Early in his career Bruce Berne and coworkers \cite{rao79} examined some of
these issues in a pioneering simulation study of the effects of confining a
Lennard-Jones (LJ) liquid-vapor system in a slit with a varying width. This
is a long-standing and very general problem, with work extending back to the
time of van der Waals, and some interesting questions remain even now.

Here we consider a simpler geometry where the different correlations
manifest themselves in a particularly clear way. We study the density
response and associated thermodynamic properties of a LJ liquid at a state
very close to the triple point as the radius of a hard sphere solute is
varied. Let $S$ denote the radius of the \emph{solute cavity} within which
the centers of fluid particles are excluded. As shown in recent computer
simulations by Huang and Chandler \cite{huang00} (HC), for small molecular
sized $S$ of order unity (with the usual LJ parameter $\sigma $ as the unit
of length) the induced density profile is highly oscillatory and resembles
the radial distribution function of the uniform LJ fluid, with a large
density maximum near the solute. However for much larger $S$ the solute
approaches a hard wall, and the density response is very different: a
``drying layer'' of low vapor density forms near the wall, and the
transition to the bulk liquid density occurs through an essentially
unperturbed and smooth vapor-liquid interface. In general, as $S$ increases
from molecular size, the contact value of the density profile decreases and
mixed states with increasingly damped oscillations are seen.

We will develop a quantitative description of this process, generalizing
ideas based on mean field theory that we have developed in a series of
recent papers \cite{wsb,wvk,wkv,kw,vkw}. Consider first the qualitative
physics. As pointed out by Widom \cite{widomsci}, in most typical
configurations in the uniform LJ fluid the attractive intermolecular forces
on a given particle tend to cancel in pairs between oppositely situated
particles. This leaves only the excluded volume correlations induced by the
harshly repulsive forces, which are well approximated by those in a uniform 
\emph{hard sphere reference system}. This cancellation argument should also
apply, though somewhat less accurately, to a small hard sphere solute of the
same molecular size in the LJ fluid. The density response to such a fixed
particle should thus closely resemble the oscillatory radial distribution in
the associated hard sphere reference model, with a density \emph{maximum} at
contact.

However, since the hard core solute provides no balancing attractive forces,
as its radius increases, LJ\ particles near the solute will increasingly
experience a net ``uncancelled'' or ``unbalanced'' attractive force directed 
\emph{away} from the solute arising from neighboring particles situated
farther away \cite{wsb,wvk}. This unbalanced attractive interaction induces
a second more slowly varying ``interfacial component'' in the density
response that competes with the oscillatory excluded volume correlations and
tends to produce a \emph{lower} density near the solute. The final density
profile results from the interplay between these components, and depends on
how close the thermodynamic state of the initial uniform LJ fluid is to
coexistence and on the magnitude of the unbalanced force, clearly related in
this case to the size of the hard core solute.

The quantitative theory behind these ideas makes use of a fundamental
approximation: the attractive interactions are described by a general \emph{%
mean} or \emph{molecular field} approach in which the locally averaged
effects of the attractive intermolecular interactions are replaced by an
effective single particle potential. Since the attractive interactions are
relatively slowly varying, it seems plausible that such an averaged
description could often provide a useful simplification. Thus, in the
present case the structure of the nonuniform LJ system is approximated by
that of a (hard sphere) reference fluid in an appropriately chosen \emph{%
effective reference field }(ERF) comprised of two parts: the bare external
field from the hard core solute and a much more slowly varying part
describing the unbalanced attractive interactions. 

Of course, there are inherent errors in the theory arising from the use of
an ERF to describe the effects of attractive interactions. In addition,
other errors can arise from the numerical determination of the properties of
the reference fluid in the presence of the ERF. However, we have previously
introduced a new and generally very accurate way of calculating the
structure and thermodynamics of the reference fluid in the presence of a 
\emph{general} external field, based on a locally optimized application of
linear response theory \cite{kw,vkw}. Thus, the latter source of error is
very small (as will be verified later in this paper) and the quantitative
validity of the theory for the LJ-hard core solute system depends mainly on
the choice of the ERF.

In our earlier studies of nonuniform fluids in a variety of different
geometries, we obtained very good qualitative and often even quantitative
accuracy \cite{wvk,wkv,vkw} using the simplest possible mean or molecular
field description of the ERF, given in Eq. (\ref{mfint}) below. However in
the present case, where an essentially unperturbed vapor-liquid interface
with a very large change in density can form at large $S$, the usual
molecular field theory will yield shifted (mean field) values for the
coexisting vapor and liquid densities. The main problem with the theory is
not so much its description of the local density gradients, but its
predictions for the thermodynamic properties of the coexisting \emph{bulk}
phases. In this paper we introduce a simple modification of the mean field
expression for the ERF that incorporates information from an accurate
equation of state for the \emph{uniform} fluid. The modified expression for
the ERF insures that \emph{exact} thermodynamic results (consistent with the
given uniform fluid equation of state) are found in the ``hydrostatic
limit'' of a very slowly varying ERF, and it gives in particular accurate
results for the coexisting vapor and liquid densities.

Using this modified expression for the ERF, we obtain excellent agreement
with the simulations of HC for the density response to the hard sphere
solute for all values of $S$ tested. HC have already shown that good
qualitative agreement can be obtained using a simplified version of the mean
field theory introduced by Lum, Chandler, and Weeks \cite{lcw} (LCW). The
LCW theory does not require detailed information about the intermolecular
potential (and hence can be applied to drying transitions in more
complicated liquids like water) but empirically fits a few key quantities
using experimental data for properties of the liquid-vapor system. The
present results suggest that the full theory can be used for quantitative
calculations as well in simple liquids, requiring only the intermolecular
potential and an accurate bulk equation of state along with known properties
of the hard sphere reference system.

\section{Simple molecular field approximation for the ERF}

Fluid particles interact with a known external field $\phi (\mathbf{r})$
from the hard core solute. We consider a grand ensemble with fixed chemical
potential $\mu ^{B}$, which determines $\rho ^{B}$, the uniform fluid
density far from the solute where $\phi (\mathbf{r})=0.$ The LJ pair
potential $w(r)\equiv u_{0}(r)+u_{1}(r)$ is separated into rapidly and
slowly varying parts associated with the intermolecular \emph{forces} \cite
{wca} so that all the harshly repulsive forces arise from $u_{0}$ and all
the attractive forces from $u_{1}$.

The structure of the nonuniform LJ system is related to that of a simpler 
\emph{nonuniform} \emph{reference fluid }\cite{wsb,wvk,wkv}, with only
repulsive intermolecular pair interactions $u_{0}(r)$ (equal to the LJ
repulsions) and a chemical potential $\mu _{0}^{B}$ corresponding to the
same bulk density $\rho ^{B},$ but in a different renormalized or \emph{%
effective reference field} $\phi _{R}(\mathbf{r})$. (In the numerical
calculations that follow, the soft-sphere reference fluid is approximated by
a hard sphere fluid of appropriately chosen diameter, with the finite
softness of $u_{0}$ taken into account by the usual ``blip function''
expansion \cite{awc,hansenmac}, as described in detail in earlier work \cite
{kw,vkw}. The errors introduced by this treatment of the reference fluid are
very small, and for most purposes the reference system can be thought of
simply as a hard sphere system.)

What is the best choice for $\phi _{R}(\mathbf{r})?$ Since we want the
reference fluid structure to approximate that of the full fluid to the
extent possible, it seems reasonable to determine $\phi _{R}(\mathbf{r})$
formally by the requirement that it produces a \emph{local} (singlet)
density at every point $\mathbf{r}$ in the reference fluid equal to that of
the full nonuniform LJ fluid \cite{sullstell}: 
\begin{equation}
\rho _{0}(\mathbf{r};[\phi _{R}])=\rho (\mathbf{r};[\phi ])\,.
\label{singletden}
\end{equation}
In practice we will make approximate choices for $\phi _{R}$ motivated by
mean field ideas. The subscript $0$ in Eq.~(\ref{singletden}) denotes the
reference fluid and the notation $\mathbf{[}\phi _{R}]$ indicates that all
distribution functions are functionals of the appropriate external field.

We can derive a formally exact equation for such a $\phi _{R}$ by
subtracting the first equations of the YBG hierarchy \cite{hansenmac} for
the full and reference systems with $\phi _{R}$ chosen so that Eq. (\ref
{singletden}) is satisfied \cite{wsb,wvk,wkv}. The result appropriately
focuses on \emph{forces} \cite{widomsci,wca} and can be written exactly as 
\begin{eqnarray}
-\nabla _{1}[\phi _{R}(\mathbf{r}_{1})-\phi (\mathbf{r}_{1})] &=&-\int d%
\mathbf{r}_{2}\rho _{0}(\mathbf{r}_{2}|\mathbf{r}_{1};[\phi _{R}])\nabla
_{1}u_{1}(r_{12})  \label{exactybg} \\
&&-\int d\mathbf{r}_{2}\{\rho (\mathbf{r}_{2}|\mathbf{r}_{1};[\phi ])-\rho
_{0}(\mathbf{r}_{2}|\mathbf{r}_{1};[\phi _{R}])\}\nabla _{1}w(r_{12})\,. 
\nonumber
\end{eqnarray}
Here $\rho _{0}(\mathbf{r}_{2}|\mathbf{r}_{1};\mathbf{[}\phi _{R}])\equiv
\rho _{0}^{(2)}(\mathbf{r}_{1},\mathbf{r}_{2}\mathbf{;[}\phi _{R}])/\rho
_{0}(\mathbf{r}_{1}\mathbf{;[}\phi _{R}])$ is the \emph{conditional} singlet
density, i.e., the density at $\mathbf{r}_{2}$ given that a particle is
fixed at $\mathbf{r}_{1}$.

To integrate this equation and obtain a simple expression for the effective
field $\phi _{R}$, we make some physically motivated approximations. If we
assume that Eq. (\ref{singletden}) produces similar local environments for
the (identical) repulsive cores in the two fluids, which then mainly
determine density correlations through excluded volume effects, then when
Eq. (\ref{singletden}) is satisfied, the conditional singlet densities in
the two fluids should also be very similar. This suggests that the last term
on the right side in Eq. (\ref{exactybg}) should be very small, at least at
high densities where excluded volume correlations dominate. Moreover the
last term clearly vanishes at very low density where $\phi _{R}=\phi .$

If we ignore the last term entirely, we obtain an approximate equation for $%
\nabla \phi _{R}$ suggested by Weeks, Selinger and Broughton \cite{wsb}. We
focus here on the even simpler \emph{molecular field equation} that arises
from further approximating the conditional singlet density $\rho _{0}(%
\mathbf{r}_{2}|\mathbf{r}_{1};\mathbf{[}\phi _{R}])$ in the first term on
the right side by the ordinary singlet density $\rho _{0}(\mathbf{r}_{2};%
\mathbf{[}\phi _{R}]).$ This approximation is much better than one might at
first suppose, since the main difference in these two functions occurs when $%
\mathbf{r}_{2}$ is close to $\mathbf{r}_{1},$ but then for small $r_{12}$
the multiplicative factor $-\nabla _{1}u_{1}(r_{12})$ (the \emph{attractive}
part of the LJ force) vanishes identically. The gradient $\nabla _{1}$ can
then be taken outside the integral and the equation can be integrated.
Choosing the constant of integration so that $\phi _{R}^{MF}$ vanishes far
from the solute where the density equals $\rho ^{B}$, we obtain the simple 
\emph{molecular field }(MF) equation \cite{wvk} for the ERF:

\begin{equation}
\phi _{R}^{MF}(\mathbf{r}_{1})-\phi (\mathbf{r}_{1})=\int d\mathbf{r}
_{2}\,\rho _{0}(\mathbf{r}_{2};\mathbf{[}\phi
_{R}^{MF}])\,u_{1}(r_{12})+2\rho ^{B}a,  \label{mfint}
\end{equation}
where 
\begin{equation}
a\equiv -\frac{1}{2}\int d\mathbf{r}_{2}\,u_{1}(r_{12})  \label{aint}
\end{equation}
corresponds to the attractive interaction parameter $a$ in the uniform fluid
van der Waals equation, as discussed below.

A self-consistent solution of Eq. (\ref{mfint}) must be found, since the ERF 
$\phi _{R}^{MF}$ appears explicitly on the left side and implicitly on the
right side through the dependence of the reference density $\rho _{0}(r_{2};%
\mathbf{[}\phi _{R}^{MF}])$ on $\phi _{R}^{MF}$. As mentioned above, we have
developed new and generally very accurate ways of calculating $\rho _{0}(r;%
\mathbf{[}\phi _{R}])$ for a given $\phi _{R}$. Since these methods have
been described in some detail previously \cite{kw,vkw}, we will not review
them here. Using these methods, it is straightforward to solve the molecular
field equation (by iteration, for example) to determine the self-consistent $%
\phi _{R}^{MF}$ and the associated density $\rho _{0}(r;\mathbf{[}\phi
_{R}^{MF}])$, which from Eq. (\ref{singletden}) approximates that of the
original LJ-hard core solute system.

While the simple molecular field expression (\ref{mfint}) often gives quite
satisfactory results \cite{wvk,wkv,vkw}, particularly for local density
gradients, for quantitative accuracy in the present application we require a
better treatment of two phase coexistence. In the limit of a \emph{uniform}
system, Eq. (\ref{mfint}) describes all effects of attractive interactions
in terms of the \emph{constant} parameter $a$ as in the van der Waals
equation. Indeed the theory then reduces to the generalized van der Waals
theory of Longuet-Higgins and Widom \cite{lhwidom}, where one combines an
accurate description of the uniform (hard sphere) reference system with the
simple treatment of the attractive interactions in terms of the constant $a.$
This very simple approximation captures much essential physics and gives a
good qualitative description of the uniform fluid thermodynamic properties.
To achieve quantitative agreement with bulk thermodynamic properties one can
replace the constant $a$ by a \emph{function} $\alpha $ that depends
(hopefully weakly, to the extent the van der Waals theory is reasonably
accurate) on temperature and density \cite{wb83}. We will adopt such a
strategy here, modifying Eq. (\ref{mfint}) so that it incorporates
information from a given accurate bulk equation of state and gives exact
results when $\phi $ and $\phi _{R}$ are very slowly varying.

While the details differ, our method is similar in spirit to the procedure
introduced in important work by van Swol and Henderson \cite{vsh91}. They
modified the usual mean field description of attractive interactions in
density functional theory to ensure that accurate coexistence behavior is
obtained in a density functional treatment of a square well fluid at a
square well wall. The main idea behind both approaches is that when
repulsive forces are accurately described, relatively simple modifications
of the van der Waals treatment of attractive forces can produce accurate
thermodynamic and structural data in the limit of very slowly varying fields
or in bulk phases. The modified theory combines the simplicity of mean field
theory with an accurate description of two phase coexistence.

\section{Constant fields and chemical potential changes}

Consider first the important special case where the external field $\phi $
is a \emph{constant} everywhere. Since in the grand ensemble, one encounters
only the difference between the chemical potential and the external field,
the effect of such a field just amounts to a change in the chemical
potential of the uniform LJ fluid from $\mu ^{B}$ to $\mu ^{B}-\phi ,$ which
then changes the bulk density \cite{hansenmac,kw}. Equation (\ref{mfint})
then reduces to the simple molecular field \emph{approximation} for the
analogous field $\phi _{R}$ (or the reference chemical potential change)
that from Eq. (\ref{singletden}) is supposed to produce the same density
change in the uniform reference fluid. In general there will be errors
arising from the molecular field approximation even in this simple limit.
However, if an accurate equation of state for the uniform fluid is known, we
can calculate these density changes exactly and this will suggest how to
modify Eq. (\ref{mfint}) so that it gives exact results for very slowly
varying fields.

To that end, let $\mu (\rho )$ denote the chemical potential as a function
of density $\rho $ for the uniform LJ fluid. (This also depends on the
temperature, but we are interested in density variations along particular
isotherms, so we will not indicate the temperature dependence explicitly.)
We determine $\mu (\rho )$ from the accurate 33-parameter equation of state 
\cite{johnson93} given by Johnson, et al. This provides a very good global
description of the stable liquid and vapor phases in the (cutoff and
shifted) LJ fluid used in the simulations of HC and provides a smooth
interpolation in between by using analytic fitting functions. Thus it
naturally produces a modified ``van der Waals loop'' in the two phase region
and seems quite appropriate for our use here in improving the simplest mean
field or van der Waals description of the uniform fluid. One could
alternately use the method of van Swol and Henderson \cite{vsh91} and
achieve essentially identical results. If $\phi (\rho )$ denotes the
magnitude of the uniform external field that yields a particular density $%
\rho $ when applied to the LJ system with chemical potential $\mu ^{B}$,
then its value is exactly given by 
\begin{equation}
\phi (\rho )=\mu ^{B}-\mu (\rho ).  \label{phirho}
\end{equation}

Using known properties of the hard sphere fluid, we also have essentially
exact expressions for $\mu _{0}(\rho ),$ the chemical potential as a
function of density in the uniform reference fluid, and for the inverse
function $\rho _{0}(\mu ),$ giving the uniform reference density as a
function of chemical potential. The analogous uniform reference field
yielding the same density $\rho $ as in Eq. (\ref{phirho}) is then exactly
given by 
\begin{equation}
\phi _{R}(\rho )=\mu _{0}^{B}-\mu _{0}(\rho ).  \label{phiRrho}
\end{equation}

Finally the exact $\mu (\rho )$ and $\mu _{0}(\rho )$ can be related through
a function $\alpha (\rho )$ defined so that 
\begin{equation}
\mu (\rho )=\mu _{0}(\rho )-2\rho \ \alpha (\rho ).  \label{mualpha}
\end{equation}
Thus the exact chemical potentials in the uniform LJ and reference systems
are related in the same way as predicted by the simple mean field
approximation of Eq. (\ref{mfint}) except that the constant $a$ in Eq. (\ref
{aint}) is replaced by a (temperature and density dependent) function $%
\alpha (\rho )$ chosen so that Eq. (\ref{mualpha}) holds. Since even the
simplest mean field theory is qualitatively accurate we expect that the
ratio $\alpha (\rho )/a$ will be of order unity and rather weakly dependent
on density and temperature.

\begin{figure}[!t]
\centerline{
\epsfxsize=4.5in
\epsfbox[60 30 680 530]{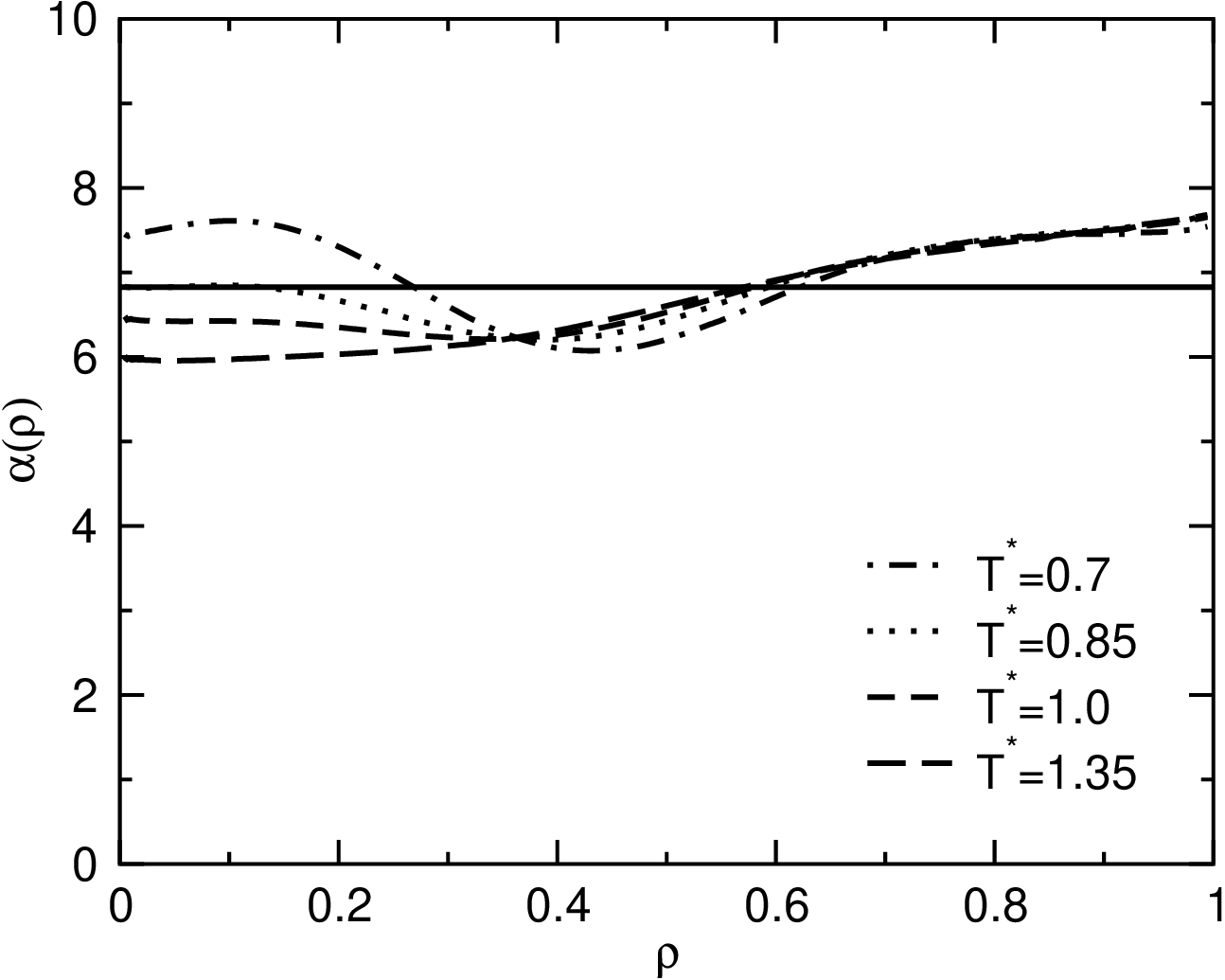}}
\caption{Density and temperature dependence of the effective
van~der~Waals parameter $\alpha(\rho)$ of the truncated and shifted LJ
fluid obtained from the accurate equation of state
\cite{johnson93}. Also shown with a solid line is the prediction of the
simple van der Waals mean field theory from Eq.~(\ref{aint}).}
\label{fig:alpha}
\end{figure}

In Fig.~\ref{fig:alpha} we give the calculated effective $\alpha (\rho )$
from the equation of state of Johnson et al. \cite{johnson93}, together with
the value $a$ of the van~der~Waals constant in Eq. (\ref{aint}) for a number
of different isotherms. The value of the constant $a=-2\pi
\int\limits_{0}^{\infty }\!dr\ r^{2}u_{1}(r)$ seems to be a good overall
compromise and may suffice for many qualitative purposes. Our derivation
resulting from approximately integrating the force balance in Eq. (\ref
{exactybg}) indicates why this expression with $u_{1}$ giving all the
attractive \emph{forces} and integrated from $0$ to infinity is appropriate.
This gives much better results than the earlier intuitive suggestion $%
a^{\prime }=-2\pi \int\limits_{\sigma }^{\infty }\!dr\ r^{2}w(r),$ which
integrates only the \emph{negative part} of the LJ potential from the point $%
\sigma $ where the potential changes sign \cite{avalues}. However the true $%
\alpha (\rho )$ exhibits some noticeable variations in density and
temperature, illustrating the need for an accurate equation of state for
quantitative accuracy. We note that some of the thermodynamic states plotted
in Fig.~\ref{fig:alpha} are in the two phase region of the bulk LJ fluid and
thus are not thermodynamically or sometimes even mechanically stable. The
analytic equation of state provides a smooth interpolation through these
states and produces behavior analogous to the van der Waals loop in the
simplest mean field description. By introducing a function $\alpha (\rho )$
that reduces to the \emph{constant} $a$ in the van der Waals theory we
hopefully have removed much of the (temperature and) density dependence
otherwise arising from the loops, leaving the relatively smooth $\alpha
(\rho )$ suitable for interpolation \cite{criticalint}.

\section{Slowly varying fields and the hydrostatic density}

These results for a strictly constant field can be used to determine simple
and accurate approximations for very slowly varying fields. In particular
when the field $\phi _{R}(\mathbf{r)}$ varies so slowly that it is
essentially \emph{constant} over the range of a correlation length in the
reference fluid, then the density $\rho _{0}(\mathbf{r}_{1};\mathbf{[}\phi
_{R}])$ at a given point $\mathbf{r}_{1}$ is very accurately approximated 
\cite{percus64,kw} by the \emph{local hydrostatic density }$\rho ^{\mathbf{r}%
_{1}}\equiv \rho _{0}(\mu _{0}^{\mathbf{r}_{1}})$, the density of the \emph{%
uniform} reference fluid at the shifted chemical potential $\mu _{0}^{%
\mathbf{r}_{1}}\equiv \mu _{0}^{B}-\phi _{R}(\mathbf{r}_{1}).$ Since in the
hydrostatic approximation the density at any given point responds only to
the \emph{local} value of the field, all the equations defined above for
constant fields and uniform densities also hold for very slowly varying
fields and the corresponding hydrostatic densities.

By subtracting Eq. (\ref{phirho}) from Eq. (\ref{phiRrho}) and using Eq. (%
\ref{mualpha}) we then arrive at an equation analogous to the simple mean
field equation (\ref{mfint}) that gives essentially exact results in the
hydrostatic limit of very slowly varying fields: 
\begin{equation}
\phi _{R}(\mathbf{r}_{1})=\phi (\mathbf{r}_{1})-2\rho ^{\mathbf{r}%
_{1}}\alpha (\rho ^{\mathbf{r}_{1}})+2\rho ^{B}\alpha (\rho ^{B}).
\label{phiRrhohydro}
\end{equation}

\section{Modified molecular field equation}

We now compare this exact result to the hydrostatic limit of the simple mean
field equation (\ref{mfint}). It is easy to see that this takes the same
form as Eq. (\ref{phiRrhohydro}) except that the function $\alpha (\rho )$
is replaced by the constant $a.$ Our goal is to modify Eq. (\ref{mfint}) so
that in the hydrostatic limit it reduces exactly to Eq. (\ref{phiRrhohydro}%
), while still giving reasonable results for more rapidly varying fields.

There is no unique way to do this, but the following simple prescription
seems most natural, and gives our final result, which we will call the \emph{%
\ modified molecular field} (MMF) approximation for the ERF: 
\begin{equation}
\phi _{R}(\mathbf{r}_{1})-\phi (\mathbf{r}_{1})=\frac{\alpha (\rho ^{\mathbf{%
r}_{1}})}{a}\int d\mathbf{r}_{2}\,\rho _{0}(\mathbf{r}_{2};\mathbf{[}\phi
_{R}])\,u_{1}(r_{12})+2\rho ^{B}\alpha (\rho ^{B}).  \label{mmfint}
\end{equation}
Thus the molecular field integral in Eq. (\ref{mfint}) is multiplied by a
factor $\alpha (\rho ^{\mathbf{r}_{1}})/a$ of order unity that depends on $%
\mathbf{r}_{1}$ through the dependence of the hydrostatic density $\rho ^{%
\mathbf{r}_{1}}$ on the local value of the field $\phi _{R}(\mathbf{r}_{1}),$
and the constant of integration $2\rho ^{B}a$ is replaced by the appropriate
limiting value of the modified integral. Note that the hydrostatic density $%
\rho ^{\mathbf{r}_{1}}$ remains smooth and relatively slowly varying even
when $\phi _{R}(\mathbf{r}_{1})$ contains a hard core. The nonlocal
oscillatory excluded volume correlations that can exist in the full density $%
\rho _{0}(\mathbf{r}_{1};\mathbf{[}\phi _{R}])\,$do not appear in $\rho ^{%
\mathbf{r}_{1}}$ because of the strictly local response to $\phi _{R}$.

The next sections report results for the structure and thermodynamics of the
the LJ-hard core solute system studied by HC with the ERF determined from
Eq. (\ref{mmfint}). We use methods based on generalized linear response
theory \cite{kw,vkw} to determine the structure of the nonuniform reference
fluid.

\section{Results}

\subsection{Structure}

To make contact with the work of HC, we study the LJ liquid at a state near
the triple point with $\rho ^{B}=0.70$ and $T=0.85$ in the presence of a
hard sphere solute. (We use the standard LJ reduced units.) By definition
the solute centered at the origin interacts with the LJ particles through
the hard core potential: 
\begin{equation}
\phi (r;S)=\left\{ 
\begin{array}{ll}
\infty , & r\leq S, \\ 
0, & r>S.
\end{array}
\right.  \label{phirS}
\end{equation}

The MMF theory discussed above allows us to reduce this problem to that of
the \emph{reference} fluid in the presence of the effective field $\phi
_{R}(r;S)$ satisfying Eq. (\ref{mmfint}). By construction from Eq. (\ref
{singletden}), the density profile of the full LJ fluid subjected to the
``bare'' external field $\phi (r;S)$ and the profile of the reference LJ
fluid in the presence of the external field $\phi _{R}(r;S)$ ``dressed'' by
the attractive interactions are supposed to be identical to each other.

\begin{figure}[t]
\centerline{
\epsfxsize=4.5in
\epsfbox[60 30 680 530]{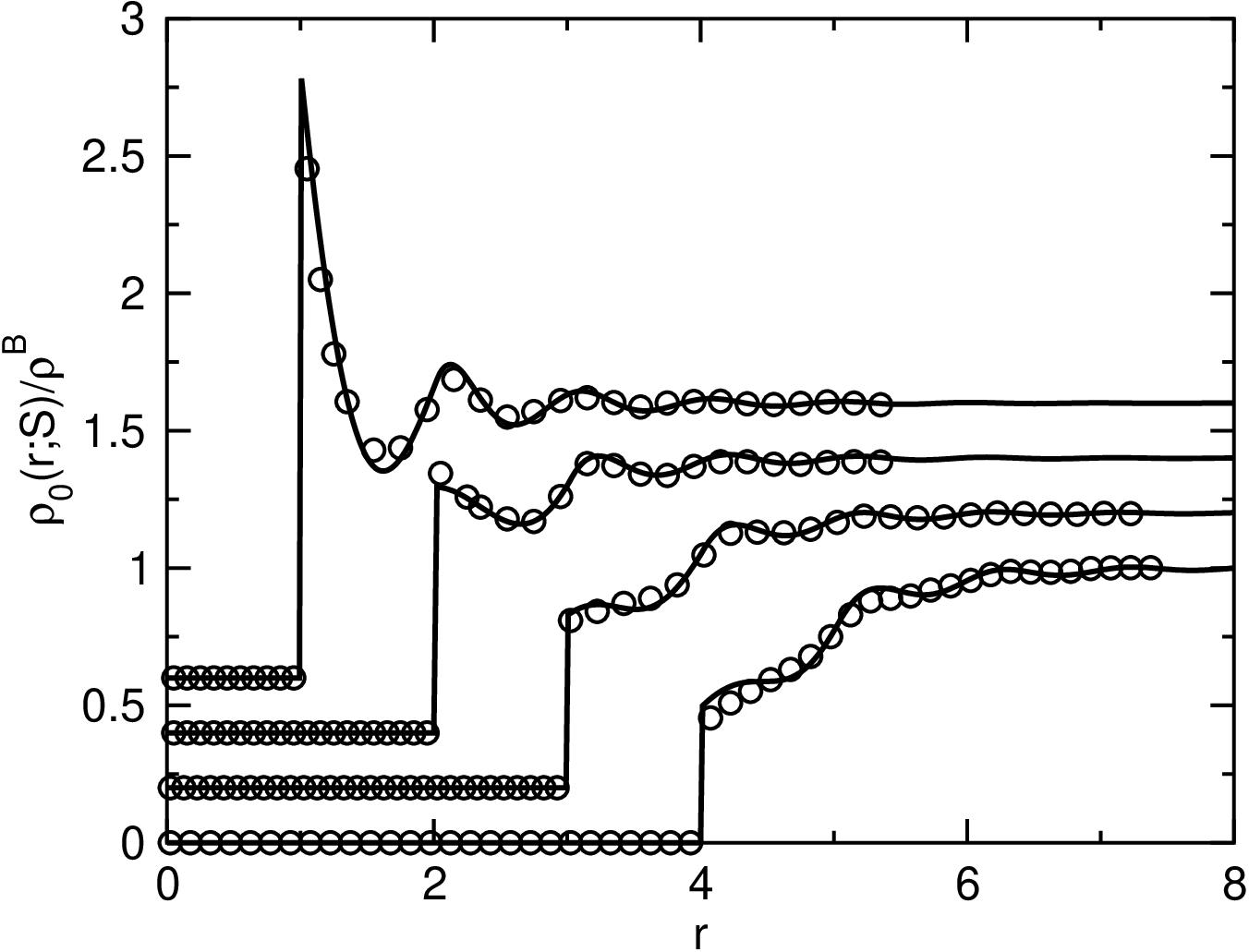}}
\caption{Density profiles of the LJ fluid ($T=0.85$, $\rho^{B}=0.70$)
in the presence of the hard sphere solute with $S=1.0$, $2.0$, $3.0$ and $4.0$.
Circles denote simulation results \cite{huang00}.
Lines are results of the self-consistent approach based on the modified mean field
determined from Eq.~(\ref{mmfint}).
For ease of viewing, the density profiles for $S=1.0$, $2.0$ and $3.0$
have been shifted vertically by $0.6$, $0.4$ and $0.2$ units respectively.}
\label{figLJhuang}    
\end{figure}

We have calculated \emph{self-consistently} the ERF $\phi _{R}(r;S)$ and the
associated density response $\rho _{0}(r;S)$ of the reference fluid, solving
Eq. (\ref{mmfint}) by iteration. In Fig.~\ref{figLJhuang} we compare these
results for the density profiles in the presence of the hard sphere solutes
with $S$ equal to $1.0$, $2.0$, $3.0$ and $4.0$ in reduced units with the
simulation results \cite{huang00} of the same LJ system by HC. There is very
good agreement between theory and simulation.

\begin{figure}[!t]
\centerline{
\epsfxsize=4.5in
\epsfbox[60 30 680 530]{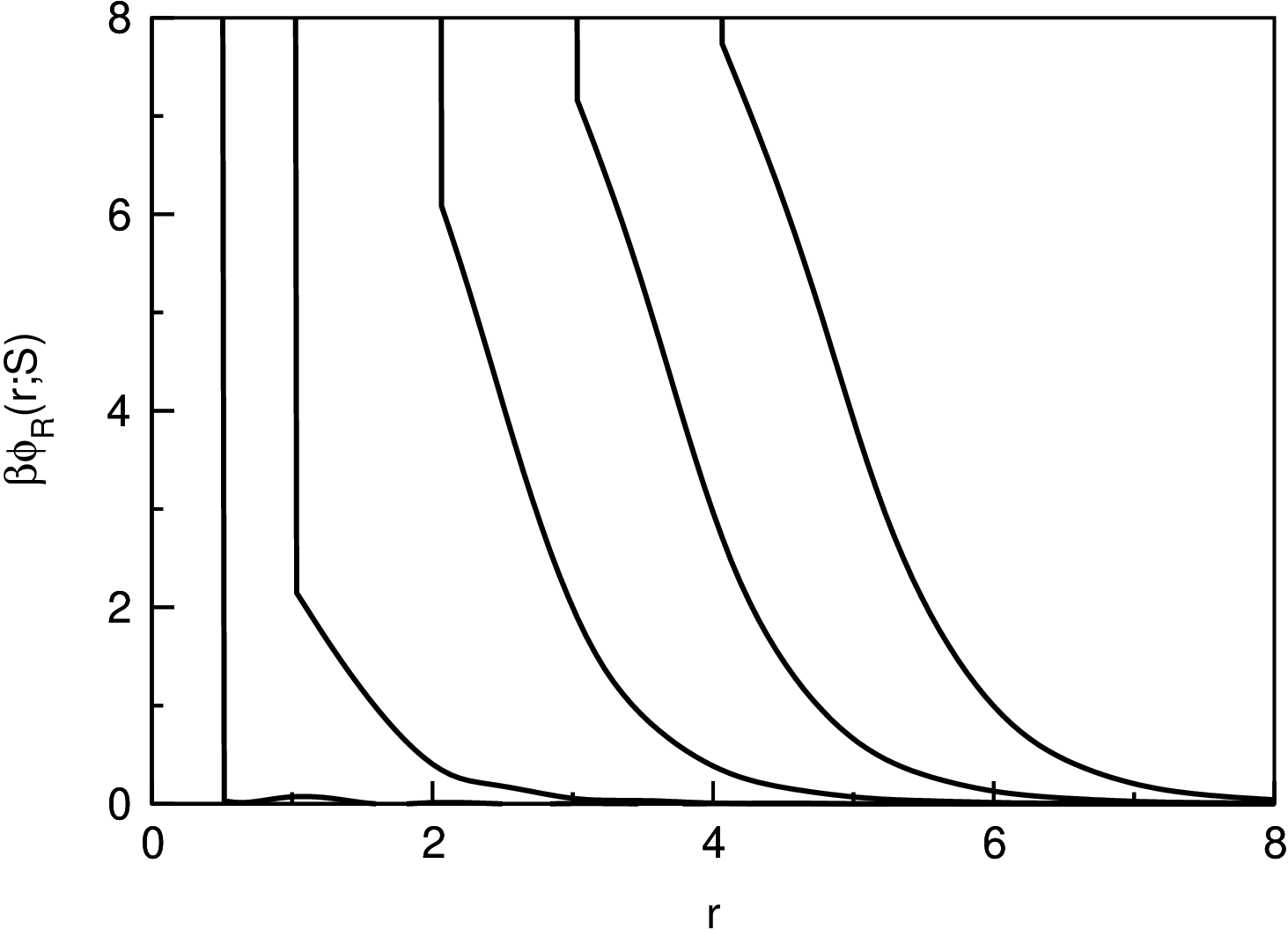}}
\caption{Self-consistent mean molecular field of the LJ fluid for the solute with  
$S=0.5$, $1.0$, $2.0$, $3.0$ and $4.0$,
obtained from  Eq.~(\ref{mmfint}).}
\label{figphiLJhuang}
\end{figure}

Figure~\ref{figphiLJhuang} shows the corresponding ERF's obtained in these
calculations. For small solutes with $S$ less than about $0.7,$ attractive
interactions do not give rise to any substantial modification of the bare
external field, as can be seen from the plot of $\phi _{R}(r;S)$ for $S=0.5$%
. (Clearly, for $S=0$ there are no solute induced iteractions of any kind
and the profile reduces to the constant $\rho ^{B}$). However, the effects
due to unbalanced attractions begin to become important even for $S=1.0$,
which is about the same size as the LJ core, and all larger sizes give rise
to a very strong and relatively soft repulsion in $\phi _{R}(r;S)$. The
corresponding density profiles show pronounced depletion near the surface of
the solute, characteristic of surface induced drying.

\begin{figure}[!t]
\centerline{
\epsfxsize=4.5in
\epsfbox[60 30 680 530]{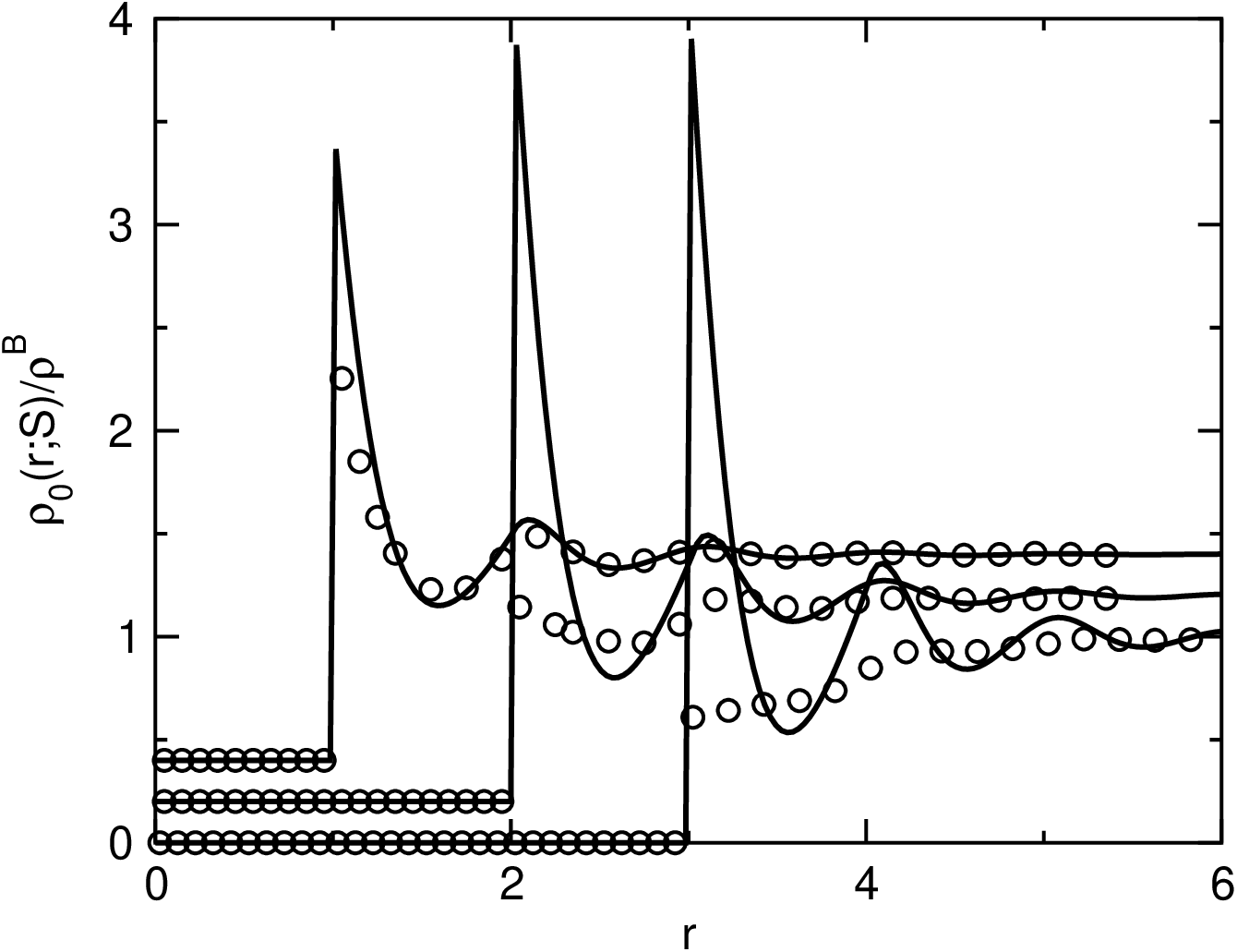}}
\caption{Density profiles of the
{\em reference} LJ fluid in the presence of the {\em bare} external field $\phi(r;S)$
for the solute with $S=1.0$, $2.0$ and $3.0$ (lines).
For ease of viewing, the density profiles for $S=1.0$ and $2.0$
have been shifted vertically by  $0.4$ and $0.2$ units respectively.
Circles denote simulation results of the {\em full} LJ fluid \cite{huang00}.}
\label{figHSref}
\end{figure}

To demonstrate the importance of the renormalization of the effective field
we show in Fig.~\ref{figHSref} the density profiles of the reference fluid
in the presence of the \emph{bare} external field $\phi (r;S)$, thus
neglecting all effects due to the attractions. These density profiles
reflect the characteristic packing effects associated with excluded volume
correlations, which result in a local ordering (or layering) of fluid
particles next to the solute. We would also expect such behavior from the LJ
system at very high temperatures, but for the state near the triple point
considered here the strong ordering is absent even though the density is
high. The unbalanced attractive interactions generate an additional strong
repulsive component in the ERF as shown in Fig.~\ref{figphiLJhuang}. This is
capable of destroying the ordering and effectively places a layer of the
metastable low density vapor next to the solute's surface for states near
coexistence.

\begin{figure}[!t]
\centerline{
\epsfxsize=4.5in
\epsfbox[60 30 680 530]{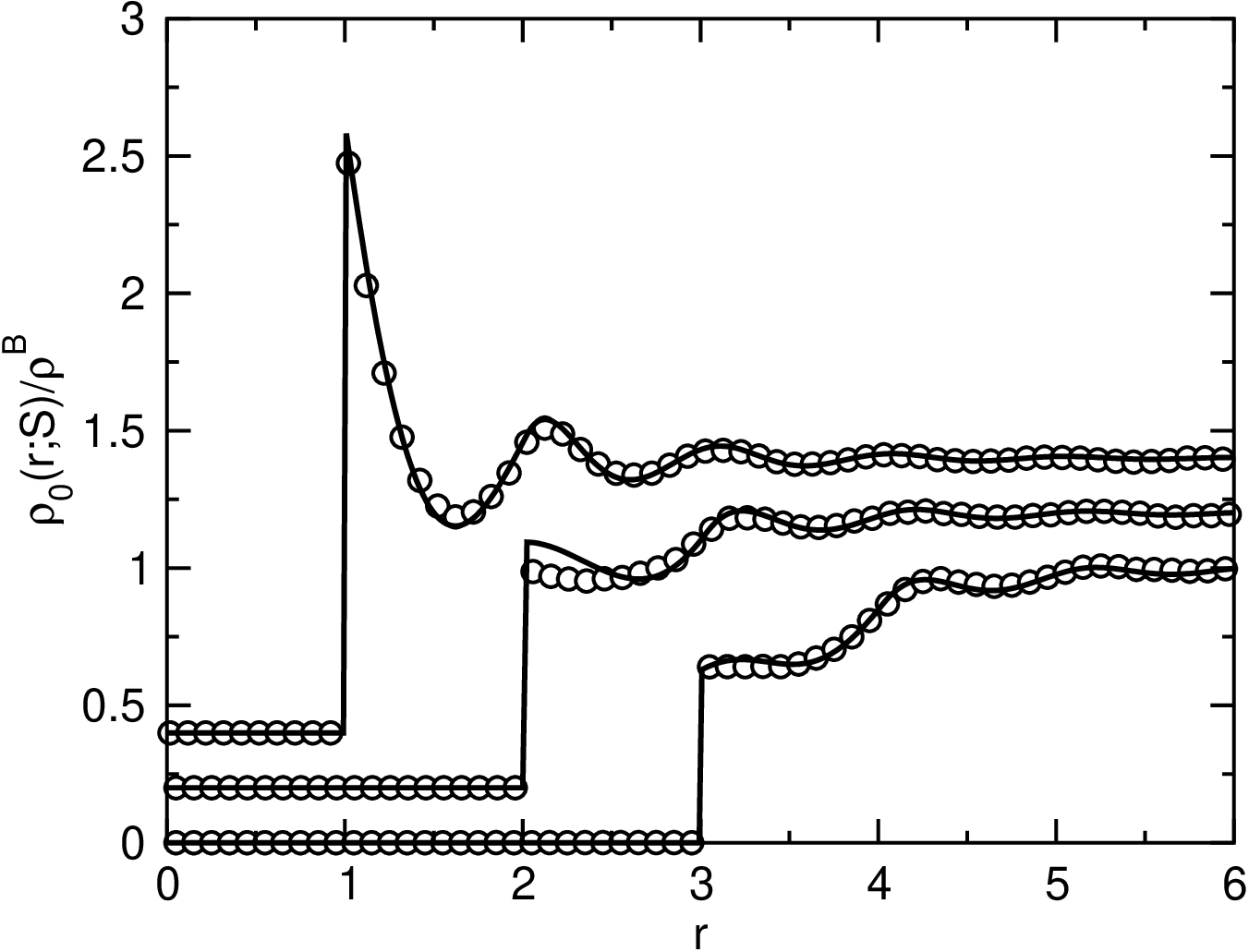}}
\caption{Density profiles of the {\em reference} LJ fluid in the
presence of the self-consistently determined mean molecular field given by
Eq.~(\ref{mmfint}) for the solute with $S=1.0$, $2.0$ and $3.0$.
For ease of viewing, the density profiles for $S=1.0$ and $2.0$
have been shifted vertically by  $0.4$ ad $0.2$ units respectively.}
\label{figLJref}
\end{figure}

In the self-consistent theory discussed above we make approximations both in
determining the ERF and in calculating the structure of the reference fluid
the presence of a given ERF. To make sure that in the present case the
agreement with the simulation results for the LJ fluid given in Fig. \ref
{figLJhuang} is not the result of some fortuitous cancellation of errors
associated with our treatment of the reference fluid, we have carried out
computer simulations \cite{kw} of the \emph{reference} fluid structure in
the presence of the self-consistently determined ERF's. This directly tests
our theory \cite{kw,vkw} for reference system correlations in the presence
of a given external field. Results of these calculations are shown in Fig.~%
\ref{figLJref} and demonstrate a very good agreement between our theory for
the reference fluid structure \cite{kw,vkw} and the simulations.

\subsection{Solvation free energy}

Another quantity of great interest is the free energy of the nonuniform
system. This is the main focus of attention in density functional theory 
\cite{evans92}, where one tries to approximate the ``intrinsic'' free energy
as a functional of the nonuniform singlet density. The equilibrium density
and value of the free energy for a given external field and set of
intermolecular interaction potentials is determined by minimizing an
appropriate free energy functional of the density and those potentials \cite
{hansenmac}. By starting from the free energy, certain exact \emph{sum rules}
relating integrals of those correlation functions to the thermodynamic
properties are automatically and consistently satisfied \cite{vsh91}. Of
course consistency does not necessarily imply correctness --- a poor
functional will give consistently poor values for the thermodynamic
properties --- and the physical implications of particular approximations
made in DFT can sometimes be hard to understand.

In contrast, our approach focuses first on the liquid \emph{structure}, as
do most integral equation methods. We believe this permits physical insight
to play a more direct role. However, since we can determine the density
response to an arbitrary external field, the free energy can be easily
calculated from a coupling parameter type integration that connects some
initial state (e.g., the bulk fluid) whose free energy is known to the final
state as the field is varied. Of course, our structural predictions, though
generally accurate, are not exact and different routes to the same
thermodynamic property can give different answers. Fortunately, in the
present case there is a very simple route to the free energy of the
nonuniform LJ system that uses structural features that we know from
simulations are accurately determined.

\subsection{Virial route}

The basis for our calculation lies in the definition of the grand canonical
free energy (thermodynamic potential) in the LJ fluid, which is related to
the partition function $\Xi $ through the following equation: 
\begin{equation}
\Omega =-k_{B}T\ln \Xi .  \label{grandfreeen}
\end{equation}
We also use the property that the partition function is the generating
function for the density correlation functions; in particular, the density
profile in the presence of a general field $\phi $ is given by 
\begin{equation}
\rho (\mathbf{r};\mu ^{B},[\phi ])=k_{B}T\frac{\delta \ln \Xi (\mu
^{B},[\phi ])}{\delta [\mu ^{B}-\phi (\mathbf{r})]}.  \label{eq:rhofromXi}
\end{equation}

We now specialize to the hard core solute $\phi (r;S)$ and introduce a
coupling-parameter dependent external field $\phi _{\lambda }(\mathbf{r}%
)\equiv \phi (\mathbf{r}/\lambda )$ with corresponding density $\rho
_{\lambda }(\mathbf{r})\equiv \rho (\mathbf{r};\mu ^{B},[\phi _{\lambda }])$%
. This $\lambda $-dependence corresponds to growing the range of the
external field from zero at $\lambda =0$ to the full external field at $%
\lambda =1$. This construction is the basis of scaled particle theory \cite
{reiss59}. As is well-known, the free energy change for the hard core field
then takes the particularly simple form: 
\begin{equation}
\beta \Delta \Omega _{S}=4\pi S^{3}\int\limits_{0}^{1}\!d\lambda \ \lambda
^{2}\rho _{\lambda }(\lambda S^{+}),  \label{HSomega}
\end{equation}
which requires only the \emph{contact value} $\rho _{\lambda }(\lambda
S^{+}) $ of the density profile. This is very accurately given by the theory
described above. To use this ``virial route'', we can discretize the $%
\lambda $-integration and calculate the density profile for several values
of $\lambda $ at the fixed bulk chemical potential $\mu ^{B}$.

\begin{figure}[!t]
\centerline{
\epsfxsize=4.5in
\epsfbox[60 30 680 530]{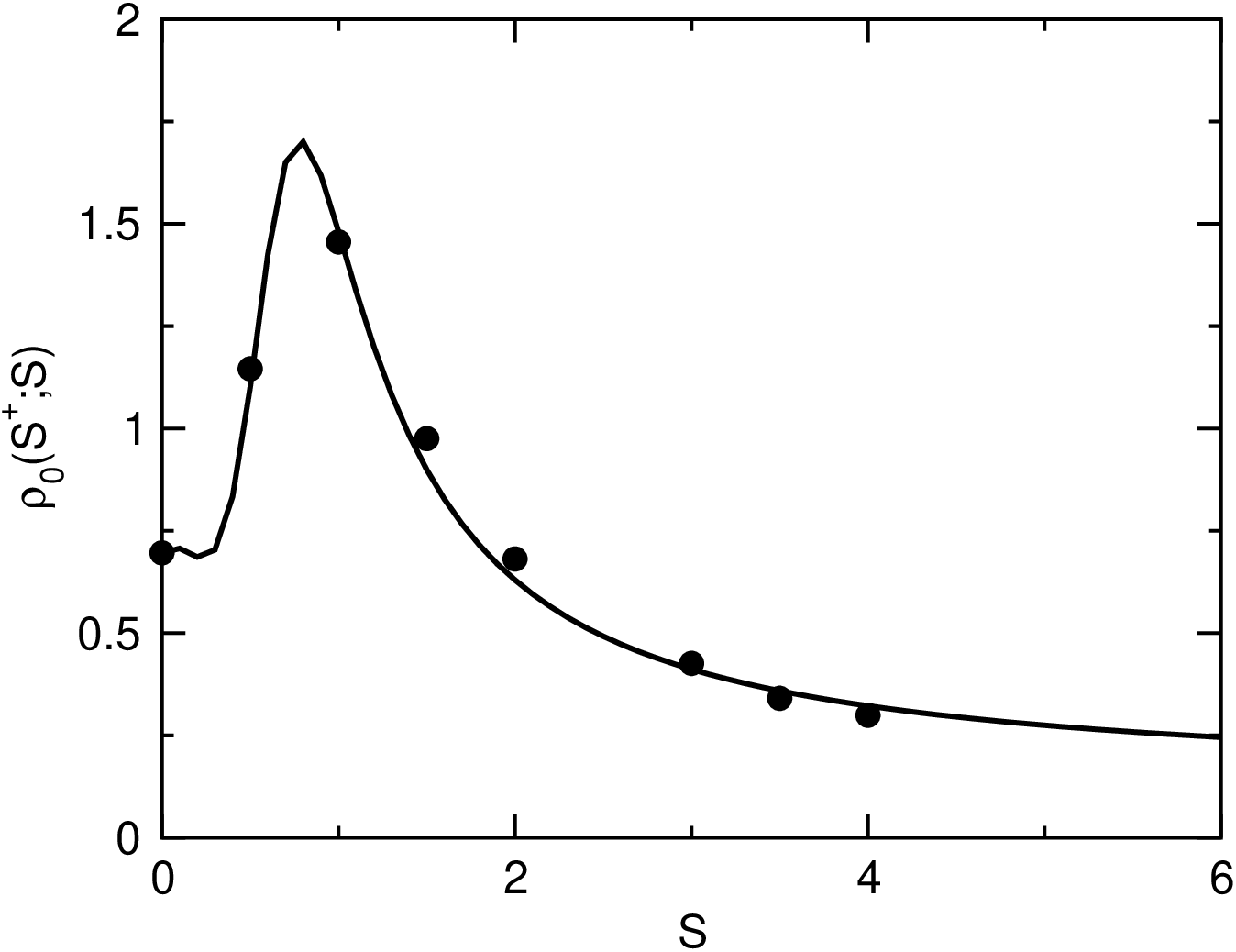}}
\caption{Dependence of the contact value of the density profiles of the LJ fluid on the
cavity size $S$. Circles denote results of simulations \cite{huang00}.
Line is the result of the present theory.}
\label{figcontact}
\end{figure}

In Fig.~\ref{figcontact} we plot the contact values as a function of the
size $S$. For solvent fluids with purely repulsive interactions, such as the
hard sphere fluid, the contact value monotonically increases because the
fluid particles increasingly tend to order next to the surface of the solute
and the density profile always has a sharp peak at contact. A nonmonotonic
dependence of the contact values is a very characteristic feature of surface
induced drying, and has been predicted for the case of water in contact with
hydrophobic objects by Stillinger \cite{stillinger73} and in the present
case by the LCW theory.

\begin{figure}[!t]
\centerline{
\epsfxsize=4.5in
\epsfbox[60 30 680 530]{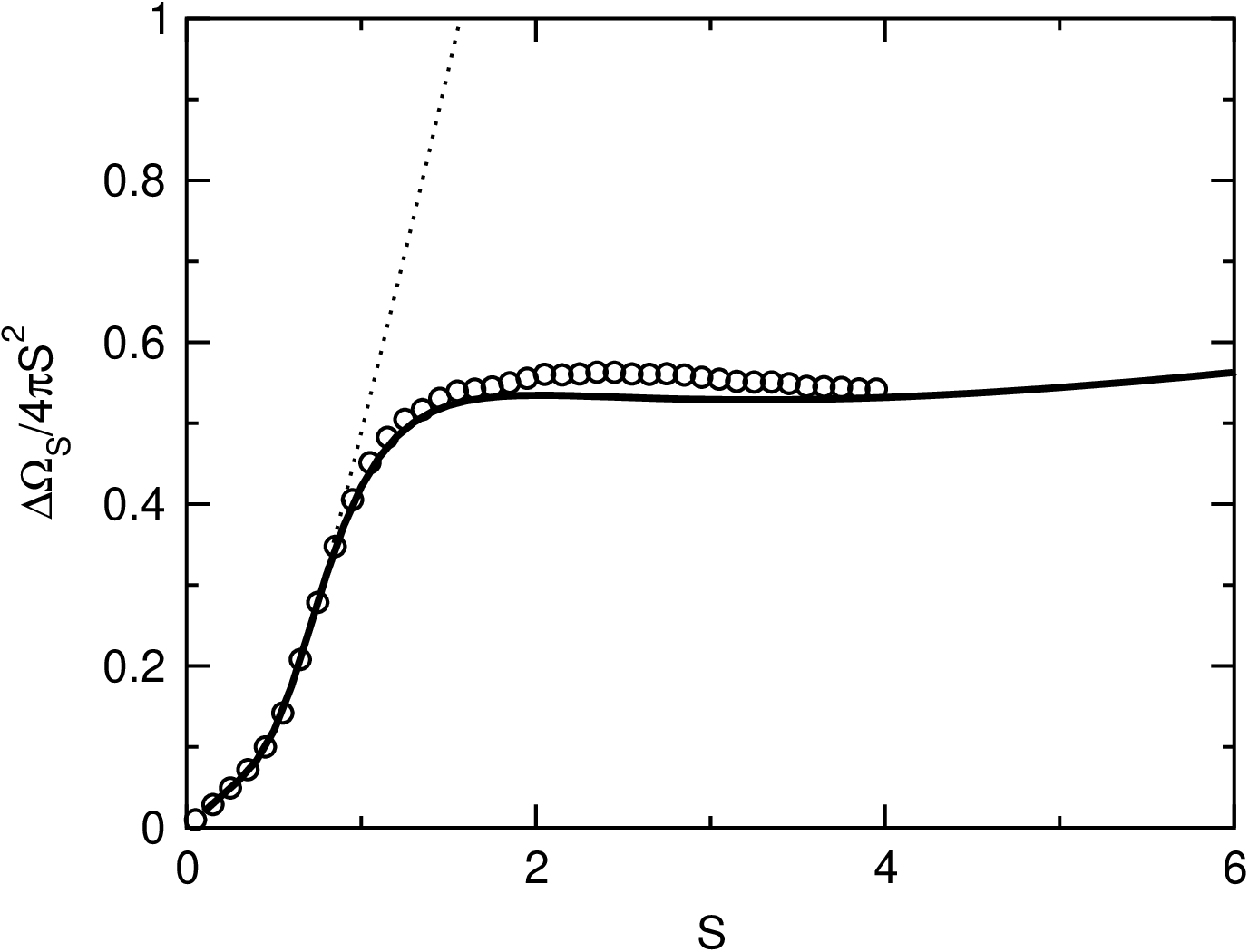}}
\caption{Dependence of the
solvation free energy on the cavity size $S$. Circles denote results of simulations
\cite{huang00}. Lines are obtained from Eq.~(\ref{HSomega}) by using the results
of the mean field Eqs.~(\ref{mmfint}) (solid), and by neglecting the mean field (dotted).}
\label{figdmu}
\end{figure}

Using Eq.~(\ref{HSomega}) we obtain the dependence of solvation free energy
on the size of the hard sphere solute. The free energy per unit surface area
of the solute $\Delta \Omega _{S}/4\pi S^{2}$ we obtain is shown in Fig.~\ref
{figdmu}. For small solutes attractive interactions do not play an important
role and the solvation free energy agrees well with a pure hard sphere
model, which completely neglects attractions by using the bare solute
potential, as shown by the dotted line. At the solute size of about $0.7$
the behavior changes drastically and the reduced free energy rapidly crosses
over to the practically constant plateau in agreement with the simulation
results. The small slope of the curves in Fig.~\ref{figdmu} for large $S$
can be understood by separating the free energy into volume ($V_{S}=4\pi
S^{3}/3$) and surface ($A_{S}=4\pi S^{2}$) contributions \cite{huang00} as
discussed by HC: 
\begin{equation}
\Delta \Omega _{S}\;\approx \;V_{S\,}p^{B}+A_{S}\gamma _{S}.
\end{equation}
The first term in this expression corresponds to the work required to remove
liquid particles from the volume occupied the solute, where $p^{B}$ is the
bulk liquid pressure, and is very small for the values of $S$ considered
here. The second term determines the cost of forming the liquid-solute
interface and is proportional to the interface tension $\gamma _{S}$, which
is essentially independent of the solute size for large solutes.

\section{Conclusions}

The theory described here combines an accurate treatment of correlations
induced in the nonuniform hard sphere reference system by a general external
field with a mean field description of the effects of attractive
interactions in terms of an appropriately chosen ERF. We have shown here how
relatively simple modification of the simple molecular field expression for
the ERF can ensure that accurate results are found in the hydrostatic limit
of very slowly varying fields and for properties of coexisting bulk phases.
Of course there are important limitations inherent in any mean field
treatment of long wavelength fluctuations, such as those seen near the
critical point or the capillary wave fluctuations of the liquid-vapor
interface \cite{hansenmac}. However in many other applications where such
fluctuations do not play an important role, mean field theory provides a
simple and often very accurate starting point. In future work we will
discuss properties of the liquid-vapor interface and of fluids confined in
slits and cylinders from this perspective.

It is a pleasure to dedicate this paper to Bruce Berne on the happy occasion
of his 60th birthday. This work was supported in part by NSF Grant No.
CHE9528915.

\end{document}